\documentclass[a4paper,12pt]{article}

\usepackage{amssymb,amsmath,empheq,epsfig,graphicx}

\numberwithin{equation}{section}

\title{\bf Asymptotic symmetries and subleading soft graviton theorem in higher dimensions}

\author{
  Dimitri Colferai
   \footnote{Email: colferai@fi.infn.it}
   \\
   {\sl\small Dipartimento di Fisica, Universit\`a di Firenze and INFN Firenze}\\
   {\sl\small Via Sansone 1, 50019 Sesto Fiorentino, Italy}
   \\[1ex]
   and
   \\[1ex]
   Stefano Lionetti
   \footnote{Email: stefano.lionetti@stud.unifi.it}
   \\
   {\sl\small Dipartimento di Fisica, Universit\`a di Firenze and INFN Firenze}\\
   {\sl\small Via Sansone 1, 50019 Sesto Fiorentino, Italy}\\[5mm]
   \\[5mm]
}

\date{}

\begin{document}

\maketitle

\begin{abstract}
We investigate the relation between the subleading soft graviton theorem and
asymptotic symmetries in gravity in even dimensions $d=2+2m$ higher than
four. After rewriting the subleading soft graviton theorem as a Ward identity,
we argue that the charges of such identity generate Diff$(S^{2m})$. In order to
show that, we propose suitable commutation relation among certain components of
the metric fields. As a result, all Diff$(S^{2m})$ transformations are
symmetries of gravitational scattering.
\end{abstract}



\section{Introduction\label{sec:intro}}

Over the last few years a triangular equivalence relation was discovered connecting three apparently different topics: asymptotic symmetries, soft theorems and memory effects. This equivalence relation can be drawn potentially in every theory with a massless particle, for example in QED, QCD, SUSY and gravity \cite{a7,a8,a9,a11,a12,a13,a14,a26,stromlect,a27,a36}. 
In this paper we examine the relationship between the subleading soft graviton theorem and asymptotic symmetries in gravity in even dimensions higher than four. 

In gravity the asymptotic symmetries are diffeomorphisms that leave invariant the asymptotic structure of spacetime transforming an asymptotically flat metric into an other asymptotically flat metric \cite{a1,a2,a3}. 
In order to define precisely asymptotic flatness, we need to specify the rate at which the metric approaches a Minkowski metric at asymptotically large distances. Unfortunately there is no unambiguous method of determining such asymptotically flat falloff conditions which are often only a posteriori justified. There are however guidelines that must be followed:
the falloff conditions should be weak enough so that all interesting solutions are allowed, but strong enough to rule out unphysical solutions, such as those with infinite energy.
Various options are discussed in the literature leading to different results. We decided to choose falloff conditions leading to non-trivial relations among $\mathcal{S}-$matrix elements.
Indeed, Ward identities, associated to proper asymptotic symmetries, turn out to be nothing but soft theorems. 

The soft graviton theorem \cite{a4,a5,a6} is a universal formula relating scattering amplitudes that differ only by the addition of a graviton whose energy $\omega$ is taken to zero
\begin{equation}
\mathcal{M}_{n+n'+1}\left(q=\omega \hat{q} ; p_{1}, \ldots, p_{n+n'}\right) = \left[S^{(1)}+S^{(2)}\right] \mathcal{M}_{n+n'}\left(p_{1}, \ldots, p_{n+n'}\right)+\mathcal{O}\left(\omega \right)
\end{equation}
where $\{p_1, \ldots p_n\}$ are the momenta of the incoming particles, while $\{p_{n+1}, \ldots p_{n+n'}\}$ are the momenta of the outgoing particles. $S^{(1)}$ and $S^{(2)}$ are given by 
\begin{equation}
\begin{aligned} S^{(1)} & \equiv \frac{\kappa}{2} \left( \sum_{k=n+1}^{n+n'} \frac{\varepsilon_{\mu \nu} p_{k}^{\mu} p_{k}^{\nu}}{p_{k} \cdot q} - \sum_{k=1}^{n} \frac{\varepsilon_{\mu \nu} p_{k}^{\mu} p_{k}^{\nu}}{p_{k} \cdot q}\right)\\  
S^{(2)} & \equiv-\frac{i \kappa}{2} \left(  \sum_{k=n+1}^{n+n'} \frac{\varepsilon_{\mu \nu} p_{k}^{\mu} q_{\rho} J_{k}^{ \rho \nu }}{p_{k} \cdot q}- \sum_{k=1}^{n} \frac{\varepsilon_{\mu \nu} p_{k}^{\mu} q_{\rho} J_{k}^{ \rho \nu }}{p_{k} \cdot q} \right) \end{aligned}
\end{equation}
where $\kappa^2=32 \pi G$ and $J_{k}^{ \rho \nu }$  is the total angular momentum (orbital+spin) of the $k$-th particle and $\varepsilon_{\mu \nu}$ is the polarization tensor of the graviton.
In the soft limit $\omega \rightarrow 0$, the leading term given by
  $S^{(1)}$ is of order $1/\omega$, while the subleading term given by
  $S^{(2)}$ is constant in $\omega$.
The formula indeed comprises the leading and subleading soft graviton
theorem. We are not concerned with higher order terms in our analysis.

In four dimensions the leading soft theorem is equivalent to the supertranslation Ward identity \cite{a7,a8}, while the subleading soft theorem is equivalent to the Diff$(S^2)$ Ward identity \cite{a15,a16}.\footnote{An alternative proposal was given in \cite{Conde:2016rom} where both the leading and subleading soft theorems were linked to supertranslations by expanding the associated charge in powers of $\frac{1}{r}$.} Together, supertranslations and Diff$(S^2)$ make up the asymptotic symmetry group. 

The equivalence relation between asymptotic symmetries and soft theorems seems to be somewhat more obscure in dimensions higher than four. Indeed soft gravitons theorems hold in any dimension $d=2m+2$ while both supertranslations and Diff$(S^{2m})$ charges seem to diverge in $d>4$ dimensions. Many researchers eliminated these divergences by imposing strong falloff conditions and leaving only the Poincar\'e group as part of the asymptotic symmetry group \cite{a19,a20,a21,a22}. 
In $d=4$, such argument would not hold.  Falloff conditions disallowing supertranslations automatically exclude all the generic radiative solutions in $d=4$.
In any case, we will not consider such strong falloff conditions in $d>4$ either, since we want to preserve the equivalence relation between asymptotic symmetries and soft theorems.
However, a renormalization seems to be mandatory in order to solve the divergence problems. We do not deal with this issue in this paper.

This paper relies on the analysis in \cite{a24} and it extends their results. In \cite{a26}  an alternative definition of the asymptotic group was proposed. However this last analysis seems to work only in the harmonic gauge while the asymptotic group in \cite{a24} can be derived in both Bondi and harmonic gauge (see appendix B in \cite{Avery:2015gxa}).

The aim of the present report is to show that the correspondence between the subleading soft theorem and asymptotic symmetries is preserved in even dimensions higher than four. 
In order to do so, we rewrite the soft theorem as a Ward identity. We then argue
that such identity is associated to Diff$(S^{2m})$ by proposing a suitable
  commutation relation among certain components of the metric fields. We expect the need for a renormalization in order to prove such commutation relation.
We work in the Bondi gauge and linearized gravity coupled to massless matter throughout the paper. Our discussion is restricted to tree-level.

The outline of the paper is as follows.
In section 2, we briefly review asymptotically flat geometries in dimensions higher than four as they were defined by Strominger et al. \cite{a24}. We then argue in favor of weaker falloff conditions in order to determine a larger group of asymptotic symmetries. This larger group is given by the semi-direct product of supertranslations and Diff$(S^{2m})$. In section 3, we rewrite the subleading soft graviton theorem as a Ward identity in the six-dimensional case. By proposing a commutation relation, we then argue that the charges we found in such Ward identity generate Diff$(S^{2m})$. In section 4, we generalize our previous results to arbitrary even dimensions higher than four.

\section{Asymptotically flat geometry}

In this section we study asymptotically flat spacetimes in $d=2m+2$ dimensions. Our attention will be limited to even dimensions due to known difficulties in defining null infinity in odd-dimensional spacetimes \cite{a39}. We will work in linearized gravity.

\subsection{Metrics} 
\begin{figure}[tbp]
\centering
\includegraphics[scale=0.4]{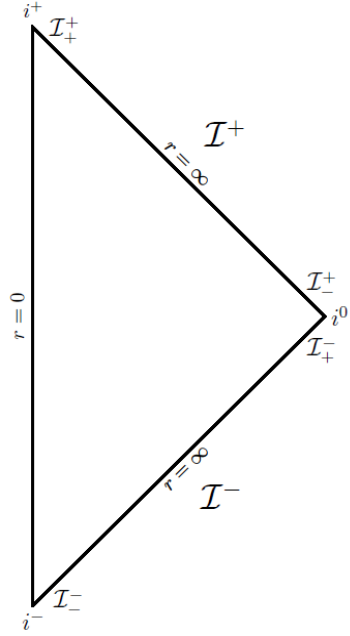}
\caption{\label{fig:penrosediagr}Penrose diagram of Minkowski space. Past and future null infinities are labelled by $\mathcal{I}^{\pm}$ and their boundaries by $\mathcal{I}^{\pm}_{\pm}$. Past and future time-like infinities are labelled by  $i^{\pm}$ and spatial infinity by $i^0$.}
\end{figure}

We are interested in asymptotically flat spacetimes at both future and past null infinity $\mathcal{I}^{\pm}$ (see figure \ref{fig:penrosediagr}).
For concreteness, let us focus on future null infinity.
We choose the coordinate system $(u,r,z^a)$, where $u=t-r$ is the retarded time, $r$ is the radial coordinate and $z^a$ ${(a=1,\dots,2m)}$ are the coordinates on the sphere $S^{2m}$. We work in the Bondi gauge imposing the following $2m+2$ conditions
\begin{equation} \label{eq:gaugefixcond}
g_{r r}=0, \quad g_{r a}=0, \quad
\operatorname{det} g_{a b}=r^{4 m} \operatorname{det} \gamma_{a b}
\end{equation}
where $\gamma_{ab}$ is the standard round metric on $S^{2m}$ with covariant derivative $D_a$. All angular indices $(a,b,c\dots)$ are raised and lowered with respect to $\gamma_{ab}$. We denote the contraction of such indices with a dot. 

According to the proposal in \cite{a24}, an asymptotically flat metric in even dimensions higher than four is given by
\begin{equation}
d s^{2}= M d u^{2}-2 d u d r+g_{a b}d z^{a}d z^{b}-2U_adz^adu
\end{equation}
where the coefficients of the metric admit an expansion near $\mathcal{I}^+$ of the form
\begin{equation} \label{eq:falloffcondmetrica}
\begin{aligned}  M=-1+\sum_{n=1}^{\infty} \frac{M^{(n)}(u, z)}{r^{n}}, \quad \quad U_{a}=\sum_{n=0}^{\infty} \frac{U_{a}^{(n)}(u, z)}{r^{n}}, \quad \quad  g_{a b}=r^{2} \gamma_{a b}+\sum_{n=-1}^{\infty} \frac{C_{a b}^{(n)}(u, z)}{r^{n}} \end{aligned}
\end{equation}
Falloff conditions on Ricci's tensor are also required
\begin{equation}
\begin{aligned} \label{eq:falloffcondricci}
R_{u u} &=\mathcal{O}\left(r^{-2 m}\right), & R_{u r}=\mathcal{O}\left(r^{-2 m-1}\right), & \quad R_{u a}=\mathcal{O}\left(r^{-2 m}\right) \\
R_{r r} &=\mathcal{O}\left(r^{-2 m-2}\right), & R_{r a}=\mathcal{O}\left(r^{-2 m-1}\right), & \quad R_{a b}=\mathcal{O}\left(r^{-2 m}\right)
\end{aligned}
\end{equation}
When the theory is coupled to matter sources, we impose the same falloff conditions on the components of $T^{M}_{\mu \nu}$ as on $R_{\mu \nu}$. In the next section we will need to slightly relax these asymptotic conditions.
The asymptotic falloff \eqref{eq:falloffcondricci} on $R_{uu}$ reads
\begin{equation} \label{eq:falloffcondricciaaa}
\frac{1}{2}\left[D^{2}+n(n+1-2 m)\right] M^{(n)}+\partial_{u} D^{a} U_{a}^{(n)}+m \partial_{u} M^{(n+1)}=0, \quad  0 \leq n \leq 2 m-3
\end{equation}
where $D^{2}=D^aD_a$. The asymptotic falloff on $R_{ur}$ reads
\begin{equation}\label{eq:falloffcondriccibbb}
\quad -\frac{n(n+1-2 m)}{2} M^{(n)}+\frac{(n-1)}{2} D^{a} U_{a}^{(n-1)}=0, \quad \quad \quad 0 \leq n \leq 2 m-2
\end{equation}
The asymptotic falloff on  $R_{ra}$ reads
\begin{equation}  \label{eq:falloffcondricciccc}
\frac{(n+2)(n+1-2 m)}{2} U_{a}^{(n)}-\frac{(n+1)}{2} D^{b} C_{b a}^{(n-1)}=0, \quad \quad  \quad 0 \leq n \leq 2 m-2
\end{equation}
Since we are working in linearized gravity, the last gauge-fixing condition in \eqref{eq:gaugefixcond} implies that all $C^{(n)}_{ab}$ are traceless
\begin{equation}
\gamma^{ab}C^{(n)}_{ab}=0
\end{equation}
So far, we have not mentioned any equations of motion. The leading $uu$ component of Einstein’s equations reads
\begin{equation}   \label{eq:eomuu}
\frac{1}{2}\left[D^{2}-2(m-1)\right] M^{(2 m-2)}+\partial_{u} D^{a} U_{a}^{(2 m-2)}+m \partial_{u} M^{(2 m-1)}+8 \pi G T_{u u}^{M(2 m)}=0
\end{equation}
The angular components of Einstein's equations determine the metric coefficients $C_{ab}^{(n)}$ recursively in terms of $C_{ab}^{(-1)}$ or $C_{ab}^{(m-2)}$ \cite{a23}
\begin{equation} \label{eq:einsteqndimrelricorsiva}
\partial_{u} D . D . C^{(n)}=\mathcal{D}_{n, m} D . D . C^{(n-1)}, \quad \quad 0 \leq n \leq m-3, \quad m-1 \leq n \leq 2 m-3
\end{equation}
with
\begin{equation}
\mathcal{D}_{n, m}=\frac{n(2 m-n-3)}{2(n+2)(-2 m+n+1)(-m+n+2)}\left(D^{2}-(n+1)(2 m-n-2)\right)
\end{equation}

\subsection{Asymptotic symmetries}

We define the asymptotic symmetry group as the group of all non-trivial diffeomorphisms preserving the asymptotic falloffs \eqref{eq:falloffcondmetrica}-\eqref{eq:falloffcondricci} and the gauge-fixing conditions \eqref{eq:gaugefixcond}. Such diffeomorphisms are generated by the vector
\begin{equation} \label{eq:vectbmsdimgentraslrot}
  \begin{aligned}
    \zeta^{u}&=f(z)+\frac{u}{2 m} D_{a} Y^{a}(z)+\ldots \\
    \zeta^{a}&=-\frac{1}{r}D^af(z)+Y^{a}(z)-\frac{u}{2mr}D^aD_bY^b(z)+\ldots \\
    \zeta^{r}&=\frac{1}{2 m}D^2f(z)-\frac{r}{2 m} D_{a} Y^{a}(z)+\frac{u}{4 m^2}D^2 D_{a} Y^{a}(z)+\ldots
  \end{aligned}
\end{equation}
where $Y^{a}(z)$ is a conformal Killing vector (CKV) on $S^{2 m}$ and $\dots$ refers to subleading terms in $r$. Transformations with $Y^{a}=0$ and an arbitrary function $f(z)$ on the sphere are known as supertranslations. 

In previous analysis \cite{a19,a20,a21,a22}, more restrictive falloff conditions were considered. That led to the conclusion that supertranslations don't exist in $d>4$.
This result seems to be at odds with the correspondence between soft theorems and asymptotic symmetries, given that the leading soft graviton theorem holds in any dimension.
The falloff conditions \eqref{eq:falloffcondmetrica}-\eqref{eq:falloffcondricci} proved to be weak enough to allow for supertranslations in dimensions higher than four. 
However, having weaker conditions, we now need to deal with divergences in computing supertranslations charges. In \cite{a25} additional boundary conditions were proposed in order to avoid these problems and still allowing for supertranslations.\footnote{These boundary conditions can be interpreted as higher-dimensional analogs of the Christodoulou-Klainerman constraints \cite{a28,a29}.} However these boundary conditions are consistent with the action of supertranslations only in linearized gravity and a renormalization is instead needed in the full theory. 

Consider now the case of $f=0$ and $Y\neq 0$ 
\begin{equation}   
\zeta =\frac{u}{2 m} D_{a} Y^{a} \partial_{u}+\Big(-\frac{r}{2 m} D_{a} Y^{a}+\frac{u}{4 m^2}D^2 D_{a} Y^{a}\Big)  \partial_{r}+\Big(Y^a-\frac{u}{2mr}D^aD_bY^b\Big) \partial_{a}+\ldots
\end{equation}
As we said, $Y$ must be a CKV on $S^{2 m}$ in order to preserve the asymptotic conditions, that is, it obeys the equation
\begin{equation}
\mathcal{L}_Y\gamma_{ab}=\frac{D.Y}{m}\gamma_{ab}
\end{equation}
In $d=4$ ($m=1$), there are infinitely many local solutions $Y$ to this last equation corresponding to infinitely many transformations generated by $\zeta$. Such transformations are called superrotations and they generalize the Lorentz transformations given by global solutions $Y$ to the last equation.
However the number of independent constraints, imposed by the CKV equation, grows with the dimensions of the space. In $d=4$ the CKV equations are the Cauchy-Riemann equations so there are infinite independent local solutions. For $d>4$, the system is over-determined and the general solution has a finite number of real parameters. 
For $d>4$, we are then left only with a finite number of independent solutions which are nothing but the global Lorentz transformations.

The asymptotic symmetry group originally defined by Bondi, van der Burg, Metzner and Sachs in $d=4$ \cite{a1,a2,a3} takes the form
\begin{equation}
\mathit{BMS}^+=\mathit{Lorentz}\ltimes \mathit{Supertranslations}
\end{equation}
Such group can then be extended by considering all local solutions $Y$ to the CKV equation as suggested by Barnich and Troessaert \cite{a40,barntro2010,barntro2011}
\begin{equation}
\mathit{Extended~BMS}^+ = \mathit{Superrotations} \ltimes \mathit{Supertranslations}
\end{equation}
The $\mathit{Extended~BMS}^+$ group preserves the asymptotic flatness locally, i.e.\ almost everywhere except in isolated points.\footnote{This violation can be physically interpreted as due to cosmic strings \cite{a37,Capone:2019aiy}.} 
In \cite{a11}, it was shown that the subleading soft graviton theorem leads to Ward identities associated to superrotations. However, it is not yet clear to what extent the superrotations symmetry implies the soft theorem. When running the argument backward, we encounter several obstacles including the need for a prescription for handling the CKV singularities. 
Whereas the Ward identities associated to supertranslations are equivalent to Weinberg’s soft graviton theorem, such equivalence could not be fully established between superrotations and the subleading soft theorem.

In order to solve this problem, a different extension of the original {\it BMS} group was proposed by Campiglia and Laddha \cite{a15,a16} in $d=4$
\begin{equation}
\mathcal{G} ^+ = \mathrm{Diff}(S^2) \ltimes \mathit{Supertranslations}
\end{equation}
General smooth vector fields on the sphere $Y$ are admitted but the assumption that $Y$ need to be a CKV is dropped.  
Therefore, we will need to relax the asymptotic falloffs \eqref{eq:falloffcondmetrica}-\eqref{eq:falloffcondricci}. 
Campiglia and Laddha shown that the Diff$(S^2)$ Ward identity is exactly equivalent to the subleading soft theorem in $d=4$.
They ensure the finiteness of the Diff$(S^2)$ charges by imposing boundary conditions.
In \cite{a18,Compere:2020lrt} the divergences are instead eliminated by adding boundary counterterms to the action. 

In dimensions higher than four, since one cannot extend Lorentz transformations to superrotations, Diff$(S^{2m})$ seem to be the proper asymptotic symmetries to link to the subleading soft graviton theorem \cite{Avery:2015gxa,Capone:2019xom}.
As we said, in order to interpret Diff$(S^{2m})$ as asymptotic symmetries, we need to consider weaker falloff conditions just like in the four-dimensional case \cite{a15,a16}. Indeed, by dropping the CKV condition, with an arbitrary smooth vector $Y$, we get
\begin{equation}
\begin{aligned}&{\mathcal{L}_{\zeta} g_{ab}=\mathcal{O}\left(r^{2}\right)} \\& {\mathcal{L}_{\zeta} g_{u u}=\mathcal{O}(1)}\end{aligned}
\end{equation}
An asymptotically flat metric should then take the form
\begin{equation}
d s^{2}= M d u^{2}-2 d u d r+g_{a b}d z^{a}d z^{b}-2U_adz^adu
\end{equation}
where the coefficients of the metric admit an expansion near $\mathcal{I}^+$ of the form
\begin{equation} \label{eq:falloffcondmetricarilassate}
\begin{aligned}  M=\sum_{n=0}^{\infty} \frac{M^{(n)}(u, z)}{r^{n}}, \quad \quad U_{a}=\sum_{n=0}^{\infty} \frac{U_{a}^{(n)}(u, z)}{r^{n}}, \quad \quad  g_{a b}=r^{2} q_{a b}+\sum_{n=-1}^{\infty} \frac{C_{a b}^{(n)}(u, z)}{r^{n}} \end{aligned}
\end{equation}
Unlike in the previous section, we don't demand $q_{ab}$ to be the round unit sphere metric and $M^{(0)}=-1$.
However we still require the falloff conditions \eqref{eq:falloffcondricci}. In particular, considering this new definition of asymptotically flat metric, the falloff condition on $R_{uu}$ implies \eqref{eq:falloffcondricciaaa} and 
\begin{equation} \label{eq:condumzero} 
\partial_u M^{(0)}=0
\end{equation}
The other equations \eqref{eq:falloffcondriccibbb}-\eqref{eq:falloffcondricciccc}, together with the Einstein equations \eqref{eq:eomuu}-\eqref{eq:einsteqndimrelricorsiva}, are unmodified.
Moreover, the $\mathcal{O}(r)$ and $\mathcal{O}(1)$ $R_{ab}$ conditions now lead
to (see app.\ \ref{app:appendixderiv2.22})
\begin{equation} \label{eq:condugammacmenuno}
  \partial_u q_{ab}=0 \;,\qquad \partial^2_u C^{(-1)}_{ab}=0
\end{equation}

We then compute the Diff$(S^{2m})$ action on the metric
\begin{equation} \label{eq:azioncmenunodiffdimgen}
\delta C_{a b}^{(-1)}=\frac{u}{2m^2}\Big[  \gamma_{a b}D^2D.Y-m\Big({D_{a} D_{b}+D_{b} D_{a}}\Big)D.Y \Big]
\end{equation}
If we consider a global Lorentz transformation, the vector $Y$ is such that $\delta C_{a b}^{(-1)}=0$.
The other angular coefficients of the metric are instead invariant under all Diff$(S^{2m})$
\begin{equation}
\delta C^{(n)}_{ab}=0 \;, \qquad {(n>-1)} \;.
\end{equation}
The action of a global Lorentz transformation is then zero on all angular components of the metric as one might expect.

There is an analogous construction on past null infinity $\mathcal{I}^-$. We impose equivalent asymptotic falloffs near $\mathcal{I}^-$ on the metric and Ricci's tensor. The metric admits an expansion in powers of $\frac{1}{r}$ near $\mathcal{I}^-$ and we denote the $n$-th expansion coefficients by $(M^{-(n)},U_a^{-(n)},C_{ab}^{-(n)})$. We also denote the asymptotic symmetry group at $\mathcal{I}^-$ by $\mathcal{G}^-$. Following the strategy firstly implemented by Strominger \cite{a7,a30}, we consider the diagonal subgroup $\mathcal{G}^0 \subset \mathcal{G}^+ \times \mathcal{G}^-$ given by the condition
\begin{equation} \label{eq:antipcondfy}
f(z)\left.\right|_{\mathcal{I}_{-}^{+}}=f(z)\left.\right|_{\mathcal{I}_{+}^{-}}, \quad \quad Y(z)\left.\right|_{\mathcal{I}_{-}^{+}}=Y(z)\left.\right|_{\mathcal{I}^{-}_{+}}
\end{equation}
which antipodally equates past and future fields near spatial infinity. Strominger showed that this antipodal matching is always possible for CK (Christodoulou-Klainerman) spaces \cite{a28,a29}.
As suggested by the reasonings in the next sections, thanks to the subleading soft graviton theorem, we argue that not only $\mathit{BMS}^0$ but the entire group $\mathcal{G}^0$ is a symmetry of gravitational scattering in arbitrary even dimensions.

\subsection{CK spaces in higher dimensions\label{s:ckhd}}

In this section we show that asymptotically flat spacetimes can
fulfill the CK constraint in higher dimensions, while allowing
Diff($S^{2m}$) to be asymptotic symmetries.  This applies to our weaker
asymptotic conditions on the metric components, which are not a source of
trouble for physical quantities at infinity. The metric itself is not
physically observable and the falloff conditions on the Ricci tensor
ensure finiteness of energy flux and other gravitational
observables. Moreover, we will show that a potentially dangerous
component of the metric is nothing but a pure “large diffeomorphism”
for CK spacetimes.

Christodoulou and Klainerman considered a class of Cauchy data decaying sufficiently fast at spatial infinity such that the Cauchy problem leads to a smooth geodesically complete solution. 
In \cite{a7} the CK conditions play a significant role in connecting $\mathcal{I}^+$ to $\mathcal{I}^-$ in $d=4$. In \cite{a24} the authors study such connection in $d>4$. As in the four-dimensional case, in order to ensure smoothness at $i^0$, the authors require higher dimensional analogs of CK constraints. Here, for the same purpose, we will impose the {\em same} CK constraints of \cite{a24}.

We consider spaces starting from the vacuum in the far past and reverting to it in the far future
\begin{equation}
\left.M^{(2 m-1)}\right|_{\mathcal{I}_{+}^{+}}=\left.M^{-(2 m-1)}\right|_{\mathcal{I}_{-}^{-}}=0,\left.\quad \quad C_{a b}^{(2 m-3)}\right|_{\mathcal{I}_{\pm}^{+}}=\left.C_{a b}^{-(2 m-3)}\right|_{\mathcal{I}_{\pm}^{-}}=0
\end{equation}
We also require that the magnetic component of the Weyl tensor vanishes near the boundaries of $\mathcal{I}^+$ and $\mathcal{I}^-$
\begin{equation}
\left.C_{u r a b}\right|_{\mathcal{I}_{\pm}^{+}}=\mathcal{O}\left(r^{-2}\right), \quad \quad \left.C_{u r a b}\right|_{\mathcal{I}_{\pm}^{-}}=\mathcal{O}\left(r^{-2}\right)
\end{equation}
Let us now focus on $\mathcal{I}^+$. 
The $\mathcal{O} (r^{-1})$ term in this constraint implies that
\begin{equation}
\lim_{u\rightarrow \pm \infty} \left(D_{a} U_{b}^{(0)}-D_{b} U_{a}^{(0)}\right)=0
\end{equation}
Using equation \eqref{eq:falloffcondricciccc} we find
\begin{equation}
\lim_{u\rightarrow \pm \infty} \left(D_{a} D^{c} C_{bc}^{(-1)}-D_{b} D^{c}C_{ac}^{(-1)}\right)=0
\end{equation}
The most general solution consistent with Bondi gauge is
\begin{equation}
C_{a b}^{(-1)}(u,z)=\frac{1}{m} \gamma_{a b} D^{2} C- \left({D_{a} D_{b}+D_{b} D_{a}}\right) C + \mathcal{O}(u^{-\epsilon})
\end{equation}
for any function $C(z,u)$ and $\epsilon>0$. 

Here is where our analysis will differ from \cite{a24}: the
asymptotic conditions $\partial_u C^{(-1)}_{ab}=0$ of \cite{a24} leads one
to consider only an angle-dependent function $C(z)$ that would not
account for Diff($S^{2m}$). Instead, our weaker asymptotic conditions%
\footnote{
These new conditions can be derived from the $R_{ab}$ asymptotic
falloff, and are still compatible with the usual CK constraints.}
$\partial^2_u C^{(-1)}_{ab}=0$ in eq.~\eqref{eq:condugammacmenuno}
admit a term linear in $u$ for $C$.
Indeed, by recalling the second equation of \eqref{eq:condugammacmenuno}, we can rewrite $C_{a b}^{(-1)}$ as pure asymptotic transformation
\begin{equation} \label{cpurasymptransf}
C_{a b}^{(-1)}(u,z)=\frac{1}{m} \gamma_{a b} D^{2} C- \left({D_{a} D_{b}+D_{b} D_{a}}\right) C
\end{equation}
with
\begin{equation} \label{eq:cpurolargediff}
\begin{aligned}
C=f(z)+\frac{u}{2m}D.Y(z)
\end{aligned}
\end{equation}
The first term of $C(u,z)$ is generated by a supertranslation while the second one is generated by a Diff($S^{2m}$). One can write an analogous equation for $C_{a b}^{-(-1)}$ on $\mathcal{I}^-$.

Finally, by taking into account that we are in CK spaces, it is always possible to require the antipodal matching condition
\begin{equation}
C_{a b}^{(-1)}|_{\mathcal{I}_{-}^{+}}=C_{a b}^{-(-1)}|_{\mathcal{I}_{+}^{-}}
\end{equation}
which in turn implies \eqref{eq:antipcondfy}.

\section{Six-dimensional gravity}

In this section we examine the relationship between Diff($S^{2m}$) and the subleading soft graviton theorem in the six-dimensional case ($m=2$).

\subsection{Mode expansions}
The fluctuation of the gravitational field in an asymptotically flat spacetime are determined by the relation $g_{\mu \nu}=\eta_{\mu \nu}+ \kappa h_{\mu \nu}$ where $\kappa^2=32\pi G$ and $\eta_{\mu \nu}$ is the flat metric. The radiative degrees of freedom of the gravitational field have the mode expansion
\begin{equation} \label{eq:modexpsei}
h_{\mu \nu}(x)=\sum_{\alpha} \int \frac{d^{5} q}{(2 \pi)^{5}} \frac{1}{2 \omega}\left[\varepsilon_{\mu \nu}^{* \alpha} a_{\alpha}(\vec{q}) e^{i q \cdot x}+\varepsilon_{\mu \nu}^{\alpha} a_{\alpha}(\vec{q})^{\dagger} e^{-i q \cdot x}\right]
\end{equation}
where $\omega=|\vec{q}|$ and $\varepsilon_{\mu \nu}^{\alpha}$ is the
polarization tensor of the graviton. The modes $a_{\alpha}$ and
$a_{\alpha}^{\dagger}$ obey the relativistic canonical commutation relations
\begin{equation}
\left[a_{\alpha}(\vec{p}), a_{\beta}(\vec{q})^{\dagger}\right]=2 \omega \delta_{\alpha \beta}(2 \pi)^{5} \delta^{5}(\vec{p}-\vec{q})
\end{equation}
In terms of the mode expansion \eqref{eq:modexpsei}, the free radiative data at $\mathcal{I}^{+}$ takes the form
\begin{equation}
C_{a b}^{(0)}(u, z) \equiv \kappa \lim _{r \rightarrow \infty} \partial_{a} x^{\mu} \partial_{b} x^{\nu} h_{\mu \nu}(u+r, r \hat{x}(z))
\end{equation}
One can evaluate the limit by a saddle-point approximation at large $r$, obtaining
\begin{equation}  \begin{aligned}
C_{a b}^{(0)}\left(u, z\right)=-\frac{2 \pi^{2} \kappa}{(2 \pi)^{5}} \partial_{a} \hat{x}^{i} \partial_{b} \hat{x}^{j} \sum_{\alpha} \int d\omega \text{ } \omega \left[\varepsilon_{i j}^{* \alpha} a_{\alpha}\left(\omega \hat{x}\right) e^{-i \omega u}+\varepsilon_{i j}^{\alpha} a_{\alpha}\left(\omega \hat{x}\right)^{\dagger} e^{i \omega u}\right] \end{aligned} 
\end{equation}
The frequency space expression is obtained by performing a Fourier transform.
The positive and negative frequency modes are then given by
\begin{equation} \label{eq:espinmodirotsei} 
\begin{aligned} C_{a b}^{\omega(0)}(z) &=-\frac{\kappa \omega}{8 \pi^{2}} \partial_{a} \hat{x}^{i}(z) \partial_{b} \hat{x}^{j}(z) \sum_{\alpha} \varepsilon_{i j}^{* \alpha} a_{\alpha}(\omega \hat{x}(z)) \\ C_{a b}^{-\omega(0)}(z) &=-\frac{\kappa \omega}{8 \pi^{2}} \partial_{a} \hat{x}^{i}(z) \partial_{b} \hat{x}^{j}(z) \sum_{\alpha} \varepsilon_{i j}^{\alpha} a_{\alpha}(\omega \hat{x}(z))^{\dagger} \end{aligned}
\end{equation}
where $\omega>0$ in both formulas. 

\subsection{Subleading soft theorem as Ward identity}
We can now introduce the following operator
\begin{equation}      
  \frac{1}{2} \lim _{\omega \rightarrow 0} \left[
    \partial_{\omega}C_{a b}^{\omega(0)}+ \partial_{-\omega} C_{a b}^{-\omega(0)}\right]
  =i \int du \text{ } C^{(0)}_{ab}u
\end{equation}
Without the derivative operators $\partial_{\pm\omega}$, the left-hand side of the previous equation would lead to one side of the leading soft theorem \cite{a24}. We claim that adding one more $\partial_{\pm \omega}$ in the definition of such operator would lead us instead to the sub-subleading soft theorem.
Plugging the frequency modes formula \eqref{eq:espinmodirotsei} into this last equation, we get
\begin{equation}   \label{eq:qwertyuio}
\begin{aligned}
&\left\langle out\left|\left(i \int du \text{ } C^{(0)}_{ab}u \right)
\mathcal{S}\right| in \right\rangle \\
&\qquad=-\frac{\kappa}{16 \pi^2} \partial_{a} \hat{x}^{i}(z) \partial_{b} \hat{x}^{j}(z)  \sum_{\alpha} \varepsilon_{i j}^{* \alpha} \text{ } \lim _{\omega \rightarrow 0} \partial_{\omega} \left\langle out\left| \omega a_{\alpha}(\omega \hat{x}) \mathcal{S}\right| in \right\rangle \\
&\qquad= -\frac{\kappa}{16 \pi^2} \partial_{a} \hat{x}^{i}(z) \partial_{b} \hat{x}^{j}(z)  \sum_{\alpha} \varepsilon_{i j}^{* \alpha} \text{ } \lim _{\omega \rightarrow 0}\left(1+\omega \partial_{\omega}\right) \left\langle out\left|  a_{\alpha}(\omega \hat{x}) \mathcal{S}\right| in \right\rangle
\end{aligned}
\end{equation}
In the last line of this equation we can recognize one side of the subleading soft theorem for an outgoing soft graviton
\begin{equation}
\lim _{\omega \rightarrow 0}\left(1+\omega \partial_{\omega}\right)\left\langle out\left|a_{\alpha}(q) \mathcal{S}\right| in \right\rangle= S_{\alpha}^{(2)}\left\langle out |\mathcal{S}| in \right\rangle
\end{equation}
where 
\begin{equation}
S_{\alpha}^{(2)}=-\frac{i\kappa}{2}\left[ \sum_{k=n+1}^{n+n'} \frac{p_{k \mu} \varepsilon_{\alpha}^{\mu \nu} q^{\lambda} J_{k \lambda \nu}}{p_{k} \cdot q}-\sum_{k=1}^{n} \frac{p_{k \mu} \varepsilon_{\alpha}^{\mu \nu} q^{\lambda} J_{k \lambda \nu}}{p_{k} \cdot q}\right]
\end{equation}
The $(1+\omega \partial_{\omega})$ prefactor on the left-hand side projects out the Weinberg pole accompanying a soft insertion.
This theorem can then be expressed in the following way
\begin{equation} \label{eq:penrihsher}
\left\langle out\left|\left(i \int du \text{ } C^{(0)}_{ab}u \right) \mathcal{S}\right| in \right\rangle=\frac{i \kappa^2}{32 \pi^2}F^{out}_{ab} \left\langle out |\mathcal{S}| in \right\rangle
\end{equation}
where
\begin{equation}\label{Foutab}
F^{out}_{ab} \equiv \partial_{a} \hat{x}^{i}(z) \partial_{b} \hat{x}^{j}(z)  \sum_{\alpha} \varepsilon_{i j}^{* \alpha} \text{ } \left[ \sum_{k=n+1}^{n+n'} \frac{p_{k \mu} \varepsilon_{\alpha}^{\mu \nu} q^{\lambda} J_{k \lambda \nu}}{p_{k} \cdot q}-\sum_{k=1}^{n} \frac{p_{k \mu} \varepsilon_{\alpha}^{\mu \nu} q^{\lambda} J_{k \lambda \nu}}{p_{k} \cdot q}\right]
\end{equation}
with $q^{\mu}=\omega\left[1, \hat{x}^{i}(z)\right]$. From equation \eqref{eq:penrihsher}, one finds
\begin{equation} 
\begin{aligned}
\frac{1}{\kappa^2}\int du \int d^4z  \sqrt{\gamma}\text{ }u D.Y (D^2-2)D^aD^b\langle out |C_{ab}^{(0)}\mathcal{S}| in \rangle = \\
\frac{1}{32 \pi^2}\int d^4z  \sqrt{\gamma}\text{ } D.Y (D^2-2)D^aD^bF^{out}_{ab}\langle out |\mathcal{S}| in \rangle
\end{aligned}
\end{equation}
where $Y$ is an arbitrary smooth vector of the sphere $S^{(4)}$.

Analogous results on $\mathcal{I}^-$ follow from the subleading soft theorem for
an incoming soft graviton. Adding together the results on $\mathcal{I}^+$ and
$\mathcal{I}^-$,  we finally have the Ward identity
\begin{equation}
\langle out |Q^+\mathcal{S}-\mathcal{S}Q^-| in \rangle =0
\end{equation}
where the charges are decomposed into a soft and a hard part
\begin{equation}
Q^{\pm}=Q_{H}^{\pm}+Q_{S}^{\pm}
\end{equation}
The soft charges are given by
\begin{equation} \label{eq:caricaseidimtrovatadateodamesoff}
\begin{aligned}
Q^+_S &= \frac{1}{\kappa^2}\int du \int d^4z  \sqrt{\gamma}\text{ }u D.Y (D^2-2)D^aD^b C_{ab}^{(0)} \\
Q^-_S &= \frac{1}{\kappa^2}\int du \int d^4z  \sqrt{\gamma}\text{ }u D.Y (D^2-2)D^aD^b C_{ab}^{-(0)}
\end{aligned}
\end{equation}
If $Y$ is a CKV, the soft charge is zero as one might expect for Lorentz transformations.\footnote{This can be proven by integrating by parts and applying all the derivatives on $Y$. Since $Y$ is a CKV, one finds $\left[D^bD^a(D^2-2)D.Y \right] \propto   \gamma_{ab}$. Given that $C^{(0)}_{ab}$ is traceless, $Q^{\pm}_S$ is then zero.}
The hard charges are given by
\begin{equation} \label{eq:caricaseidimtrovatadateodamedura} 
\begin{aligned}
& Q^+_H = -\frac{1}{16 \pi^2} \int d^4z  \sqrt{\gamma}\text{ } D.Y (D^2-2)D^aD^b \left[  \partial_{a} \hat{x}^{i}(z) \partial_{b} \hat{x}^{j}(z)  \sum_{\alpha} \varepsilon_{i j}^{* \alpha} \text{ }  \sum_{k=n+1}^{n+n'} \frac{p_{k \mu} \varepsilon_{\alpha}^{\mu \nu} q^{\lambda} J_{k \lambda \nu}}{p_{k} \cdot q}\right] \\
& Q^-_H =- \frac{1}{16 \pi^2} \int d^4z  \sqrt{\gamma}\text{ } D.Y (D^2-2)D^aD^b \left[  \partial_{a} \hat{x}^{i}(z) \partial_{b} \hat{x}^{j}(z)  \sum_{\alpha} \varepsilon_{i j}^{* \alpha} \text{ }  \sum_{k=1}^{n} \frac{p_{k \mu} \varepsilon_{\alpha}^{\mu \nu} q^{\lambda} J_{k \lambda \nu}}{p_{k} \cdot q}\right]
\end{aligned} 
\end{equation}
Let us denote the momentum of the external massless particles with $p^{\mu}_k=E_k\big(1,\hat{x}(z_k)\big)$.
In analogy with the results in the four-dimensional case, after computing the
derivatives we claim that for scalar particles one gets
\begin{equation}
\begin{aligned} \left\langle out \right| & Q_{H}^{+}\propto    i \sum_{k=n+1}^{n+n'}& \left(Y^{a}\left(z_{k}\right) \partial_{z^a_{k}}-\frac{E_{k}}{4} D_{a} Y^{a}\left(z_{k}\right) \partial_{E_{k}}\right)\left\langle out \right| \end{aligned}
\end{equation}
which represent the action of a Diff$(S^{4})$ on each outgoing particle.\footnote{See appendix \ref{app:appendixhc} for a partial proof of this equation in arbitrary dimensions.}  

\subsection{Diff$(S^{4})$ charges}

We should now proceed to show that $Q=Q_{H}+Q_{S}$ generates Diff$(S^{4})$.
However computing the Diff$(S^{4})$ charges with the covariant phase
space formalism \cite{a31,ash2,ash3,ash4,a32,a33,a34,a35} leads us to divergences. 
In fact, in order to allow for the existence of supertranslations and Diff$(S^{2m})$, we considered a phase space whose symplectic structure is divergent. This prevents us from computing a well-defined charge. The same problem would arise even if we only admitted supertranslations in the asymptotic group \cite{a25}.
One possible
solution may be to add boundary counterterms to the action in order to
cancel the divergences just like it was suggested for the case of supertranslations
in $d>4$ in \cite{a25}. 
Indeed in \cite{a18} the authors performed a successful renormalization for the Diff($S^2$) charges in $d=4$.
By choosing suitable counterterms, one should be able to derive a finite symplectic form in $d>4$ too. Using such symplectic form, one could then compute the asymptotic charges.

Here instead we follow a different path. Similarly to \cite{a24}, we use a commutation relation in order to prove that $Q$ generates Diff$(S^{4})$. We do not prove this commutation relation but one should be able to do so by using the renormalized symplectic form we mentioned.

We start by writing the following formula
\begin{equation} \label{eq:reldicommutastrominger}
\left[M^{(3)}(z)|_{\mathcal{I}^+_-}, C \left(u',z^{\prime}\right)\right]=4 \pi i G \frac{\delta^{4}\left(z-z^{\prime}\right)}{\sqrt{\gamma}}
\end{equation}
This commutation relation was postulated in \cite{a24} ---
in connection with the analogous zero-mode bracket in
  QED~\cite{Kapec:2014zla} --- but in such analysis the authors
considered only an angle-dependent function $C(z')$. They used
this formula in order to prove the equivalence between the leading
soft graviton theorem and supertranslations Ward identity.  Here, as
we discussed in section 2.3, we allow $C$ to be $u$-dependent and we
assume that the commutation relation is preserved in this case.

We will now use equation \eqref{eq:reldicommutastrominger} in order to
prove the equivalence between the subleading soft graviton theorem and
Diff$(S^{4})$ Ward identity. First of all, by taking into account
eq.~\eqref{cpurasymptransf}, we rewrite the commutation
relation~\eqref{eq:reldicommutastrominger} in the form
\begin{equation}  \label{eq:miareldicommutaisiea}
 \partial_u \left[M^{(3)}(u,z), C_{a b}^{(-1)} \left(u^{\prime}, z^{\prime}\right)\right]=4 \pi i G \left(2 D_{a} D_{b}- \frac{1}{2} \gamma_{a b} D^{2}  \right) \frac{\delta^{4}\left(z-z^{\prime}\right) \delta ( u-u^{\prime} ) }{\sqrt{\gamma}}
\end{equation}

The $\partial_u M^{(3)}$ term in the commutator can then be obtained
from the constraints on the metric applied to the $uu$-component of
Einstein's equation at the leading order:
\begin{equation}
 \partial_{u} M^{(3)}=-4 \pi G T_{u u}^{M(4)}-\frac{1}{2} \partial_{u} D^{a} U_{a}^{(2)}- \frac{1}{4}\left[D^{2}-2\right] M^{(2)}
\end{equation}
We can ignore the first term on the right-hand side since it does not contribute to the commutation relation. The second term on the right-hand side is zero thanks to the following equations
\begin{equation}
U_{a}^{(2)}=-\frac{3}{4} D^{b} C_{b a}^{(1)}, \quad \partial_u D.D.C^{(1)}=0
\end{equation}
In order to rewrite the third term, consider the following equations
\begin{equation}
M^{(2)}=-\frac{1}{2} D^{a} U_{a}^{(1)}, \quad U_{a}^{(1)}=-\frac{1}{3} D^{b} C_{b a}^{(0)}
\end{equation}
Finally, thanks to the commutation relation \eqref{eq:miareldicommutaisiea}, one gets
\begin{equation} \begin{aligned}
\frac{1}{24} \left[(D^2-2)D.D.C^{(0)} (u,z), \right.&\left.  C_{a b}^{(-1)} \left(u^{\prime}, z^{\prime}\right)\right] \\
=4 \pi i G \Big( \frac{1}{2} & \gamma_{a b} D^{2} -2 D_{a} D_{b} \Big) \frac{\delta^{4}\left(z-z^{\prime}\right) \delta ( u-u^{\prime} ) }{\sqrt{\gamma}}
\end{aligned}
\end{equation}
Given the soft charge \eqref{eq:caricaseidimtrovatadateodamesoff}, $Q^+$ indeed generates Diff$(S^{4})$:
\begin{equation}
[Q^+,C_{a b}^{(-1)}] \propto i \delta C_{a b}^{(-1)}
\end{equation}
where $\delta C_{a b}^{(-1)}$ is given by equation \eqref{eq:azioncmenunodiffdimgen}.

\section{Generalization to arbitrary even-dimensional spacetime}

We now generalize the results of the previous section to arbitrary even dimensions ${d=2m+2}$ higher than four. The plane wave expansion is given by
\begin{equation}
h_{\mu \nu}(x)=\sum_{\alpha} \int \frac{d^{2 m+1} q}{(2 \pi)^{2 m+1}} \frac{1}{2 \omega}\left[\varepsilon_{\mu \nu}^{* \alpha}(\vec{q}) a_{\alpha}(\vec{q}) e^{i q \cdot x}+\varepsilon_{\mu \nu}^{\alpha}(\vec{q}) a_{\alpha}(\vec{q})^{\dagger} e^{-i q \cdot x}\right]
\end{equation}
The positive and negative frequency modes take the form
\begin{equation} \label{eq:espinmodirotdimgen}
\begin{aligned} C_{a b}^{\omega(m-2)}(z)=& \frac{(-i)^{m} \omega^{m-1} \kappa}{2(2 \pi)^{m}} \partial_{a} \hat{x}^{j}(z) \partial_{b} \hat{x}^{k}(z) \sum_{\alpha} \varepsilon_{j k}^{* \alpha} a_{\alpha}(\omega \hat{x}(z)) \\ C_{a b}^{-\omega(m-2)}(z)=& \frac{i^{m} \omega^{m-1} \kappa}{2(2 \pi)^{m}} \partial_{a} \hat{x}^{j}(z) \partial_{b} \hat{x}^{k}(z) \sum_{\alpha} \varepsilon_{j k}^{\alpha} a_{\alpha}(\omega \hat{x}(z))^{\dagger}\end{aligned}
\end{equation}
where $\omega>0$ in both formulas. We now introduce the following operator
\begin{equation} \begin{aligned}
\frac{1}{2}\lim _{\omega \rightarrow 0}  \partial_{\omega} \left[(i \omega)^{2-m}\left(C_{a b}^{\omega(m-2)}+\right. \right. & \left. \left. (-1)^{m+1} C_{a b}^{-\omega(m-2)}\right)\right]  =(-1)^m i \int{du} \text{ } u \text{ } I^{(m-2)}\left(C_{a b}^{(m-2)}\right) 
\end{aligned}
\end{equation}
which is obtained by using Fourier transform properties. $I^{(m-2)}$ is the operator that integrates $m-2$ times with respect to $u$. 
Plugging the frequency modes formula \eqref{eq:espinmodirotdimgen} into this last equation, we get
\begin{align}
&\left\langle out\left|\left( (-1)^m i \int{du} \text{ } u\text{ }
  I^{(m-2)}\left(C_{a b}^{(m-2)}\right)  \right)
  \mathcal{S}\right| in \right\rangle \nonumber \\
  &\qquad =-\frac{(-1)^m \kappa}{4 (2\pi)^m} \partial_{a} \hat{x}^{i}(z)
  \partial_{b} \hat{x}^{j}(z)  \sum_{\alpha} \varepsilon_{i j}^{* \alpha} \;
  \lim _{\omega \rightarrow 0} \partial_{\omega}
  \left\langle out\left| \omega a_{\alpha}(\omega \hat{x})
  \mathcal{S}\right| in \right\rangle \nonumber \\
  &\qquad =-\frac{(-1)^m \kappa}{4 (2\pi)^m} \partial_{a} \hat{x}^{i}(z)
  \partial_{b} \hat{x}^{j}(z)  \sum_{\alpha} \varepsilon_{i j}^{* \alpha} \;
  \lim _{\omega \rightarrow 0}\left(1+\omega \partial_{\omega}\right)
  \left\langle out\left|  a_{\alpha}(\omega \hat{x})
  \mathcal{S}\right| in \right\rangle \label{eq:qqqqwertyuio}  
\end{align}
In the last line of this equation we can recognize one side of the subleading soft theorem for an outgoing soft graviton
\begin{equation}
\lim _{\omega \rightarrow 0}\left(1+\omega \partial_{\omega}\right)\left\langle out\left|a_{\alpha}(q) \mathcal{S}\right| in \right\rangle= S_{\alpha}^{(2)}\left\langle out |\mathcal{S}| in \right\rangle
\end{equation}
where
\begin{equation}
S_{\alpha}^{(2)}=-\frac{i\kappa}{2}\left[ \sum_{k=n+1}^{n+n'} \frac{p_{k \mu} \varepsilon_{\alpha}^{\mu \nu} q^{\lambda} J_{k \lambda \nu}}{p_{k} \cdot q}-\sum_{k=1}^{n} \frac{p_{k \mu} \varepsilon_{\alpha}^{\mu \nu} q^{\lambda} J_{k \lambda \nu}}{p_{k} \cdot q}\right]
\end{equation}
This theorem can then be expressed in the following way
\begin{equation} \label{eq:qqqqpenrihsher}
\left\langle out\left| (-1)^m \int{du} \text{ } u \text{ }  I^{(m-2)}\left(C_{a b}^{(m-2)}\right)  \text{ } \mathcal{S}\right| in \right\rangle=\frac{(-1)^m  \kappa^2}{8 (2\pi)^m}F^{out}_{ab} \left\langle out |\mathcal{S}| in \right\rangle
\end{equation}
where $F^{out}_{ab}$ is formally equal to its 6-dimensional counterpart of
  eq.~\eqref{Foutab}. 
We can now write
\begin{equation} 
\begin{aligned}
 \langle out |&   \frac{(-1)^m}{\kappa^2}  \int du \int d^{2m}z  \sqrt{\gamma}\text{ }u D.Y  \times \\&  \prod_{l=m+1}^{2m-1} \left(D^{2}-(l-1)(2 m-l) \right)   I^{(m-2)}\left( D^aD^bC_{a b}^{(m-2)} \right)  \mathcal{S}| in \rangle = \\ &
\frac{(-1)^m}{8 (2\pi)^m}\int d^{2m} z  \sqrt{\gamma}  \text{ } D.Y \prod_{l=m+1}^{2m-1} \left(D^{2}-(l-1)(2 m-l) \right)    D^aD^bF^{out}_{ab}\langle out |\mathcal{S}| in \rangle 
\end{aligned} 
\end{equation}
where $Y$ is an arbitrary smooth vector of the sphere $S^{(2m)}$. 

Analogous results on $\mathcal{I}^-$ follow from the subleading soft theorem for an incoming soft graviton. Adding together the results on $\mathcal{I}^+$ and $\mathcal{I}^-$,  we finally have the Ward identity
\begin{equation}
\langle out |Q^+\mathcal{S}-\mathcal{S}Q^-| in \rangle  =0
\end{equation}
where the charges are decomposed into a soft and a hard part
\begin{equation}
Q^{\pm}=Q_{H}^{\pm}+Q_{S}^{\pm}
\end{equation}
The soft charges are given by 
\begin{equation}
\begin{aligned}
Q^+_S =& \frac{(-1)^m}{\kappa^2} \int du \int d^{2m}z  \sqrt{\gamma}\text{ }u D.Y  \prod_{l=m+1}^{2m-1} \left(D^{2}-(l-1)(2 m-l) \right)  I^{(m-2)}\left( D^aD^bC_{a b}^{(m-2)} \right)\\
Q^-_S =& \frac{1}{\kappa^2} \int du \int d^{2m}z  \sqrt{\gamma}\text{ }u D.Y   \prod_{l=m+1}^{2m-1} \left(D^{2}-(l-1)(2 m-l) \right)   I^{(m-2)}\left( D^aD^bC_{a b}^{-(m-2)} \right)
\end{aligned}
\end{equation}
If $Y$ is a CKV, the soft charge is zero as one might expect for Lorentz transformations.
The hard charges are given by
\begin{equation}   \label{eq:caricaduradimgenteosofficdasvilupppp}
\begin{aligned}
 Q^+_H =-& \frac{(-1)^m}{4 (2\pi)^m} \int d^{2m} z  \sqrt{\gamma}\text{ } D.Y\times \\
&\prod_{l=m+1}^{2m-1} \left(D^{2}-(l-1)(2 m-l) \right)   D^aD^b \left[  \partial_{a} \hat{x}^{i}(z) \partial_{b} \hat{x}^{j}(z)  \sum_{\alpha} \varepsilon_{i j}^{* \alpha} \text{ }  \sum_{k=n+1}^{n+n'} \frac{p_{k \mu} \varepsilon_{\alpha}^{\mu \nu} q^{\lambda} J_{k \lambda \nu}}{p_{k} \cdot q}\right] 
\end{aligned} 
\end{equation}
and
\begin{equation}  
\begin{aligned}
 Q^-_H =-& \frac{(-1)^m}{4 (2\pi)^m}\int d^{2m} z  \sqrt{\gamma}\text{ } D.Y \times \\
& \prod_{l=m+1}^{2m-1} \left(D^{2}-(l-1)(2 m-l) \right)    D^aD^b \left[  \partial_{a} \hat{x}^{i}(z) \partial_{b} \hat{x}^{j}(z)  \sum_{\alpha} \varepsilon_{i j}^{* \alpha} \text{ }  \sum_{k=1}^{n} \frac{p_{k \mu} \varepsilon_{\alpha}^{\mu \nu} q^{\lambda} J_{k \lambda \nu}}{p_{k} \cdot q}\right]
\end{aligned} 
\end{equation}
In analogy with the results in the four-dimensional case, after computing the derivatives we claim that for scalar particles one gets
\begin{equation} \label{eq:qharddateosofficeaaaa}
\begin{aligned} \left\langle out \right|  Q_{H}^{+} \propto    i \sum_{k=n+1}^{n+n'} \left(Y^{a}\left(z_{k}\right) \partial_{z^a_{k}}-\frac{E_{k}}{2m} D_{a} Y^{a}\left(z_{k}\right) \partial_{E_{k}}\right)\left\langle out \right| \end{aligned}
\end{equation}
which represent the action of a Diff$(S^{2m})$ on each outgoing particle.\footnote{See appendix \ref{app:appendixhc} for a partial proof of this equation.}  

We should now proceed to show that $Q=Q_{H}+Q_{S}$ generates Diff$(S^{2m})$. 
However computing the Diff$(S^{2m})$ charges with the covariant phase space formalism leads us to divergences. As we have already said, a possible solution may be to add boundary counterterms to the action to cancel the divergences. 

Similarly to our six-dimensional analysis, we follow a different path. We propose the commutation relation 
\begin{equation}
\left[M^{(2m-1)}(z)|_{\mathcal{I}^+_-}, C \left(u',z^{\prime}\right)\right]=\frac{8 \pi i G}{m} \frac{\delta^{2m}\left(z-z^{\prime}\right)}{\sqrt{\gamma}}
\end{equation}
Using this formula and proceeding in a manner similar to section 3.3, one can prove that $Q$ generates Diff$(S^{2m})$. Note that this commutation relation is consistent with the four-dimensional case \cite{a8}.

\section{Conclusions and outlook} 

In this paper we examined the symmetry group that preserve asymptotic
  flatness of even-dimensional spacetimes.  If one assumes that such
asymptotic symmetry group contains all diffeomorphisms on the sphere
(Diff($S^{2m}$)), one needs to consider less restrictive falloff conditions than
those otherwise presented in the literature.  Such choice of weaker falloff
conditions is motivated by the correspondence between soft theorems and
asymptotic symmetries.  Indeed, starting from the subleading soft graviton
theorem in even dimensions higher than four, we derived a Ward identity.  We
then argued that such identity is associated to asymptotic symmetries with
  respect to Diff($S^{2m}$) transformations, provided a suitable commutation
  relation between metric fields holds.

As a result, Diff($S^{2m}$) are symmetries of gravitational scattering.  However
we have not tackled the divergence problem in Diff($S^{2m}$) charges, which need
a renormalization. This issue needs further investigation.  The aim of this
paper is to consolidate the correspondence between soft theorems and asymptotic
symmetries in arbitrary dimensions.  It would be worthwhile to also consider the
odd-dimensional case which we haven't dealt with due to problems with the
conformal definition of null infinity.

There are still many other open questions.  For example it would be interesting
to extend our analysis to the full nonlinear theory. One could also compute
quantum corrections at one loop since our analysis is carried out at tree level.
Lastly, the sub-subleading soft graviton theorem has been recently linked to
asymptotic symmetries in $d=4$ by Campiglia and Laddha \cite{a17}. It would be
worthwhile to extend such analysis to arbitrary dimensions.

\section{Acknowledgments}
We thank Domenico Seminara for valuable comments during the
preparation of the manuscript.
  
\section{Note added after publication}

During the revision of our paper for publication in Phys.~Rev.~D, we derived
eq.~\eqref{eq:condugammacmenuno} and we added section~\ref{s:ckhd}. Accordingly,
we updated the submission on the archive (v2 $\to$ v3). Meanwhile,
asymptotic symmetries were investigated in higher dimensions for any spin in
\cite{Campoleoni:2020ejn}. We recently became aware that our
eq.~\eqref{eq:condugammacmenuno} and \eqref{cpurasymptransf} were reported in
such paper as well.

In addition it is worth noting that, after our paper was accepted for
publication, ref.~\cite{Capone:2021ouo} appeared, where Diff($S^{2m}$) were
investigated in non-linear gravity for both even and odd dimensions.
  
\appendix

\section{Derivation of the equations \eqref{eq:condugammacmenuno}\label{app:appendixderiv2.22}}
In this section we derive the equations \eqref{eq:condugammacmenuno}.
They both follow from the falloff condition $R_{a b}=\mathcal{O}\left(r^{-2 m}\right)$.
Writing the Ricci tensor component $R_{a b}$ in the Bondi coordinates, one gets
\begin{equation}
\begin{aligned}
  R_{a b} =&-\frac{1}{r}\left(2 r \partial_{r}+2m-4\right) \partial_u{g}_{a b}
  -\frac{1}{r}\left\{\left(r \partial_{r}+2m-2\right) D_{(a} U_{b) }+2 \gamma_{a b} D . U\right\} \\
  &+\frac{1}{r^{2}}\left\{\left(D^2+r^{2} \partial_{r}^{2}
  +(2m-4) r \partial_{r}-4(m-1)\right) g_{ab }-D_{(a} D^c g_{b)c}\right\} \\
  &-2 \gamma_{ab}\left(r \partial_{r}+2m-1\right) M \\
  = &-2mr \partial_u q_{ab}
  -2 q_{ab}  
  + 2(1-m)\partial_u C^{(-1)}_{ab} + 2(1-2m) \gamma_{ab} M^{(0)} +\mathcal{O}\left(r^{-1}\right)\;.
\end{aligned}
\end{equation}
The $\mathcal{O}(r)$ component of $R_{a b}$ must vanish. This leads immediately
to the equation
\begin{equation}\label{duq}
\partial_{u} q_{a b}=0 \;.
\end{equation}
The $\mathcal{O}(1)$ component of $R_{a b}$ must vanish as well. By applying to
it the partial derivative $\partial_u$, the term with $q_{ab}$ disappear, thanks
to eq.~\eqref{duq}. Finally, by using eq.~\eqref{eq:condumzero}, we obtain
\begin{equation}
\partial_{u}^{2} C_{a b}^{(-1)}=0 \;.
\end{equation}

\section{The hard charge\label{app:appendixhc}}

Here we partially prove equation \eqref{eq:qharddateosofficeaaaa} starting from the following hard charges
\begin{equation} \label{eq:hardchargedipartenzaappendix}
\begin{aligned}
 Q^+_H =-& \frac{(-1)^m}{4 (2\pi)^m} \int d^{2m} z  \sqrt{\gamma}\text{ } D.Y
\prod_{l=m+1}^{2m-1} \left(D^{2}-(l-1)(2 m-l) \right)   D^aD^b \mathcal{F}_{ab}
\end{aligned} 
\end{equation}
where
\begin{equation}
\mathcal{F}_{ab}= \left[  \partial_{a} \hat{x}^{i}(z) \partial_{b} \hat{x}^{j}(z)  \sum_{\alpha} \varepsilon_{i j}^{* \alpha} \text{ }  \sum_{k=n+1}^{n+n'}  \frac{p_{k \mu} \varepsilon_{\alpha}^{\mu \nu} q^{\lambda} J_{k \lambda \nu}}{p_{k} \cdot q}\right] 
\end{equation}
with $q=\omega \hat x(z)$.
Let us consider the completeness relation for polarization tensors
\begin{equation}
2 \sum_{\alpha} \varepsilon_{\alpha}^{* i j}(\vec{q}) \varepsilon_{\alpha}^{k l}(\vec{q})=\pi^{i k} \pi^{j l}+\pi^{i l} \pi^{j k}-\frac{1}{2} \pi^{i j} \pi^{k l}, \quad \quad \pi^{i j}=\delta^{i j}-\frac{q^{i} q^{j}}{\vec{q}^{2}}
\end{equation}
We can then write
\begin{equation}
\begin{aligned} 
& \mathcal{F}_{ab} \propto \sum_{k=n+1}^{n+n'}  \left[ \frac{1}{4} (\partial_a \hat{x} \cdot \partial_b \hat{x}) (p_k \cdot \frac{\partial}{\partial p_k})
\right. \\ & \qquad \left. 
+\frac{(\partial_a \hat{x} \cdot p_k)(\partial_b \hat{x} \cdot p_k)}{(\hat{x} \cdot p_k)} (\hat{x} \cdot \frac{\partial}{\partial p_k})- (\partial_a \hat{x} \cdot p_k)(\partial_b \hat{x} \cdot \frac{\partial}{\partial p_k}) \right]+ (a\leftrightarrow b)
\end{aligned}
\end{equation}
Recalling the parametrization $p_k=E_k \hat{x}_k(z)$, we now focus on the term proportional to $\partial_{E_k}$  
\begin{equation} \label{eq:matcalfabdee}
\mathcal{F}_{ab} \propto \sum_{k=n+1}^{n+n'}  E_k \left( \frac{1}{4} (\partial_a \hat{x} \cdot \partial_b \hat{x}) +\partial_a \mathcal{P} \partial_b \log (1-\mathcal{P}) \right)\partial_{E_k}
+ (\dots)\partial_{z_k}
\end{equation}
with 
\begin{equation}
\mathcal{P}=\sum_{i}\hat{x}^i(z) \hat{x}^i_k(z_k)
\end{equation}
where the sum is on the spatial components of the vectors.
We now need to compute the sequence of covariant derivates in \eqref{eq:hardchargedipartenzaappendix}.
One can notice that the first term on the right-hand side of \eqref{eq:matcalfabdee} can be rewritten using $\gamma_{ab}= \partial_a \hat{x} \cdot \partial_b \hat{x}$. Such term vanishes when we apply any spherical covariant derivate to $\mathcal{F}_{ab}$.
Using the following equation \cite{a24}
\begin{equation}
\begin{aligned}
(-1)^{m} \sqrt{\gamma} \prod_{l=m+1}^{2 m-1}\left[D^{2}-(2 m-l)(l-1)\right] D^{a} D^{b} \left(\partial_a\mathcal{P} \partial_b \log (1-\mathcal{P}) \right)  \\
=(2 m-1) \Gamma(m) 2^{m}(2 \pi)^{m}\left[\sum_{k} \delta^{2 m}\left(z-z_{k}\right)\right]
\end{aligned}
\end{equation}
one can then see that the term proportional to $\partial_{E_k}$ in \eqref{eq:qharddateosofficeaaaa} is correctly reproduced.


\end{document}